# Very fast optical flaring from a possible new Galactic magnetar

A. Stefanescu[1], G. Kanbach[1], A. Słowikowska[2,3], J. Greiner[1], S. McBreen[1], G. Sala[1]

[1] Max-Planck-Institute for Extraterrestrial Physics, PO Box 1312, 85741 Garching, Germany

[2] IESL, Foundation for Research and Technology - Hellas, PO Box 1385, GR-711 10 Heraklion, Greece

[3] Copernicus Astronomical Center, Rabiańska 8, 87-100 Toruń, Poland

**Highly luminous rapid flares are characteristic of processes around compact objects like white dwarfs, neutron stars or black holes. In the high energy regime of X- and γ-rays, outbursts with variability time-scales of seconds and faster are routinely observed, e.g. in γ-ray bursts[1] or Soft Gamma Repeaters[2]. In the optical, flaring activity on such time-scales has never been observed outside the prompt phase of GRBs[3]. This is mostly due to the fact that outbursts with strong, fast flaring usually are discovered in the high-energy regime. Most optical follow-up observations of such transients employ instruments with integration times exceeding tens of seconds, which are therefore unable to resolve fast variability. Here we show the observation of extremely bright and rapid optical flaring in the galactic transient[4,5,6,7] SWIFT J195509.6+261406. Flaring of this kind has never previously been reported. Our optical light-curves are phenomenologically similar to high energy light-curves of Soft Gamma Repeaters and Anomalous X-ray Pulsars[8], which are thought to be neutron stars with extremely high magnetic fields (magnetars). This suggests similar emission processes may be at work, but in contrast to the other known magnetars with strong emission in the optical.**

At 20:52:26 UT on June 10, 2007, the Burst Alert Telescope (BAT)[9] aboard the Swift Satellite[10] triggered on a possible GRB of approximately 5s duration[11]. Subsequent observations in the X-rays (by XRT[12] and Chandra[13]) and in the optical bands have shown a point source compatible with the BAT error-circle. Although a chance coincidence between the original BAT-trigger and the object discussed here is possible due to the number of potential sources in the 1.8' (90% confidence level) BAT error-circle, the 1.0" uncertainty (90% confidence level) of the XRT position is small enough to exclude a chance coincidence between the X-ray and optical bands. The optical[4,5,6,14] and X-ray afterglow behaviour indicates that the source was most probably not a GRB, but rather a galactic X-ray transient[5]. It was therefore assigned the name SWIFT J195509.6+261406[15] (hereafter SWIFT J1955).

Optical observations began just 421s after the BAT trigger with OPTIMA-Burst[16]. A total of ~8.5h of data was obtained in the first five nights after the burst (Fig. 1). The overall optical light-curve shows two brief flares shortly after the initial high-energy trigger on the first night[4] (2007-06-10 in Fig. 1). The second night shows a marked increase in flaring activity, culminating in two extremely bright flares with complex sub-structure (2007-06-11 in Fig. 1)[14]. The morphology of the light-curve, isolated brief bursts consisting of just one short peak and a bunching of these bursts in spurts of activity, is reminiscent of major X-ray outbursts in SGRs[17].

The two most prominent flares, recorded near the end of the build-up in activity on 2007-06-11, are shown in detail in Fig. 2. A notable feature of these light-curves is the extremely steep rise, followed by a slower exponential decay. Both flares are very bright, with peak I-band magnitudes of 14.8 mag and 14.2 mag respectively. Superimposed on both flares are a number of secondary flares, again with a distinct fast-rise exponential-decay (FRED) shape. From our data we can deduce an upper limit for the quiescent emission of 20 mag. This means the source changed brightness by a factor of over 200 in only 4 seconds. Kasliwal et al.[7] detected a quiescent counterpart at an I-band magnitude of 24.4±0.2 mag. Assuming the same quiescent emission level during our observation, the source brightened by a factor of more than $10^4$.

The high intrinsic time-resolution of OPTIMA-Burst (4 μs) and the signal to noise ratio of the brightest parts of the two flares means that we can resolve the light-curve to a time resolution of ~10ms. This enables us to fit a simple model of several superimposed FRED curves to the light-curve segments shown in fig.2. The shortest time-scales found in this analysis are 0.3-0.4 s. This places a limit on the maximum size of the emitting region of ~$10^{10}$ cm (~0.1 solar radii), because the light-travel time across a larger region would blur the observed features.

In order to judge the luminosity and estimate the total energy output in the optical of the flares shown in Fig. 2, information about the source distance is required. A detailed discussion of distance estimates to SWIFT J1955 based on the X-ray extinction and optical, near infra-red and radio observations is presented by Castro-Tirado et al.[18]. The lower limit to the distance as estimated by Castro-Tirado et al. is 2-4 kpc, the most probable distance lies in the range of 4-8 kpc. In this work, we adopt a reference distance of 5 kpc.

Assuming a distance d, the flares have a maximum extinction-corrected (assumed $A_V$=5) I-band isotropic luminosity of $1.0*10^{35}$ $(d/5kpc)^2$ erg/s (for the flare displayed in Fig. 2a) and $1.8*10^{35}$ $(d/5kpc)^2$ erg/s (Fig. 2b), and a total emitted energy in the I-band of $4.6*10^{36}$ $(d/5kpc)^2$ erg (Fig. 2a) and $3.2*10^{36}$ $(d/5kpc)^2$ erg (Fig. 2b). Taking into account the aforementioned size of the emitting region and putting the source at a distance of 5 kpc, a black-body temperature of several $10^7$ K is necessary to explain the observed I-band flux with thermal radiation. However, the extreme contrast ratio between quiescent emission and peak brightness and the very short rise and decay times observed in these transitions make it more probable that a non-thermal process is the source of the observed flares.

One possible non-thermal scenario is a magnetar, similar to an AXP or SGR. Optical and near IR emission has been observed previously in AXPs[19,20] , but not with strong or rapid variability. So far, no optical/NIR emission has been observed in any known SGR. Considering the similarities between the optical light-curve observed in SWIFT J1955, and X-ray light-curves of SGR outbursts[17], it is quite conceivable that this is the first observation of optical flares in a magnetar. It is surprising that the strong X-ray flaring activity usually observed in such a source was not detected in this case. There were no X-/γ-ray observations simultaneous with our or any other reported optical observations, so the amount of correlated X-ray/optical activity can only be assessed indirectly. The optical flaring activity during the first and second night of observations (2 flares/h and >7 flares/h respectively) contrasts significantly with the X-ray activity detected by XRT[7,18] between these two epochs. During the 5.5h XRT observed, 10-40 flares would be expected if a strict correlation between optical and X-ray emission pertains (which is not necessarily the case, as in the case of the AXP XTE J1810-197, reported by Testa et al.[21]) and the activity of the source were the same. However, only one significant flare of 35s duration was observed by XRT.

Since most of the observational data on magnetars so far has been in the high energy regime, the optical emission of magnetars was somewhat neglected in theory. One intriguing model proposing optical ion cyclotron emission was published by Beloborodov and Thompson[22]. In this model, coherent microwave and radio emission emitted near the neutron star is absorbed higher in the magnetosphere by ions at their cyclotron resonance, and then re-emitted in the optical nearer to the poles where the ion cyclotron cooling and transit times become comparable.

The unique high time resolution of the OPTIMA-Burst data enable us to analyse the variability of the source using the power spectral density of the light-curve (PSD). The PSD of the two very bright flares (Fig. 2) are shown in Fig. 3. The power spectrum remains flat above a few Hz up to 1 kHz, which confirms very low variability on time-scales shorter than ~0.1s. At several frequencies, both PSDs show features similar to what in the X-ray regime is called quasi-periodic oscillations (QPOs). Most notably, prominent features at a frequency of 0.16±0.03 Hz (panel a) and 0.12±0.03Hz (panel b) overlap within their full width at half maximum (FWHM). These features also coincide with a suggestion of possible X-ray periodicity reported by Kasliwal et al. at a frequency of 0.1446 Hz.[7] Interestingly, the period of 6-8 s implied by these observations lies in the range of typical periods[8,2] for AXPs and SGRs. If this is indeed the rotational period of the magnetar, the radius of its light-cylinder is about $3*10^{10}$ cm. This is consistent with the assumption of magnetospheric emission, since the upper limit for the size of the emitting region is smaller than the light-cylinder.

Two groups have proposed a connection between SWIFT J1955 and the black hole X-ray binary V4641 Sgr[15,7], in one case proposing a new class of objects[7]. However, the optical properties described here are quite different from those reported for V4641 Sgr. The light-curve of SWIFT J1955 is dominated by extremely bright (up to more than 5 magnitudes) and very fast (few seconds to sub-second) flares, whereas the optical variability of V4641 Sgr is much less extreme. The light-curves of V4641 Sgr contain flares of ~10 min duration and amplitudes of 0.05-0.2 mag during outbursts. Superimposed upon the slower variations are optical flashes of ~50s length and ~0.5mag amplitude[23,24]. Even taking into account that these values were derived from CCD-based observations with integration times of 30s to minutes, it is clear that SWIFT J1955 is much more extreme both in magnitude of the variability, as well as in the time-scales involved.

It is clear that such optical observations of high-energy transients at high time resolution can enhance our understanding of compact objects. In the case of SWIFT J1955, such observations, independently confirmed by Castro-Tirado et al.[18] with multi-wavelength observations, indicate a connection between this transient and magnetars.

Acknowledgments:

We thank the Skinakas Observatory for their support and allocation of telescope time, and acknowledge the allocation of Chandra DDT time (ObsId 8562). We thank F. Schrey, T. Kougentakis and G. Paterakis for technical support, A. de Ugarte Postigo for access to private data taken simultaneous to some of our observations and A. Castro-Tirado for discussions. Skinakas Observatory is a collaborative project of the University of Crete, the Foundation for Research and Technology - Hellas, and the Max-Planck-Institute for Extraterrestrial Physics.

A. St. acknowledges support from OPTICON

A. Sł. acknowledges support from the EU

S.M.B. acknowledges the support of the European Union through a Marie Curie

Intra–European Fellowship within the Sixth Framework Program.

G.S. is supported through DLR

Correspondence and requests for material should be addressed to A. St (astefan@mpe.mpg.de) or G. K. (gok@mpe.mpg.de)

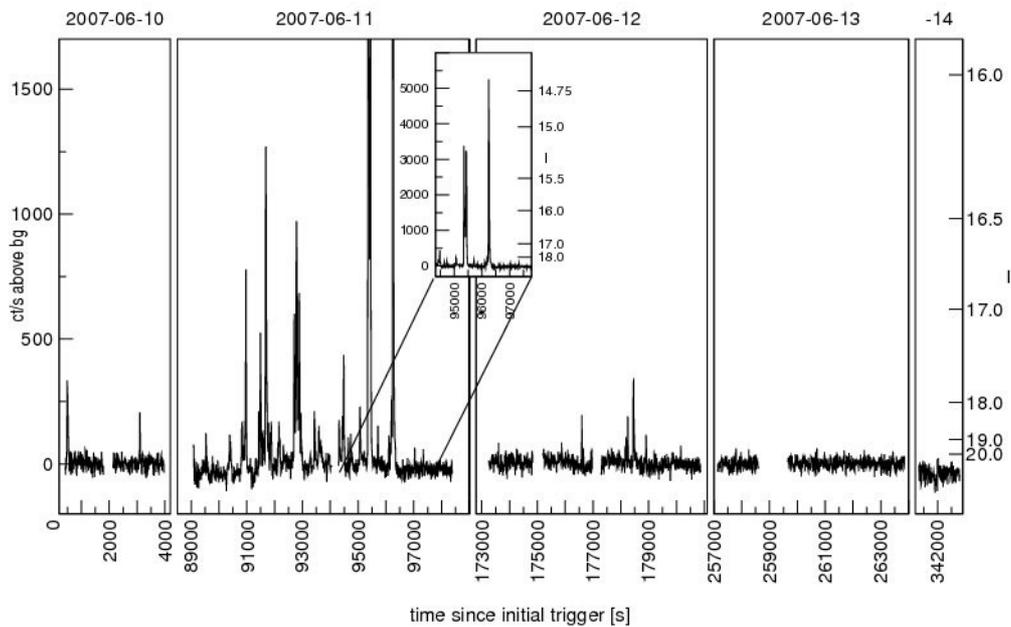

Figure 1: An overview of all optical high-time-resolution light-curves of SWIFT J1955 obtained with OPTIMA-Burst mounted at the 1.3 m Telescope of the Skinakas Observatory, Crete. OPTIMA-Burst is a fibre-fed system using six fibre apertures in a hexagonal bundle around the target fibre plus one additional more distant background fibre to determine and subtract the sky background. All apertures are of 6" diameter. The photon counting mode of OPTIMA-Burst has an intrinsic photon arrival time resolution of 4 μs.

 To achieve a high signal-to-noise ratio, the recorded photon arrival times are binned into 10 s bins in the overview light-curve shown here. The two most prominent flares at 95400 s and 96250 s post-trigger are too bright to be shown adequately next to the rest of the flares, and are therefore displayed in a different scaling in the inset. Due to the very crowded field of SWIFT J1955, most of the available background channels were severely contaminated by field stars. The three least polluted background channels were used for background subtraction, but due to changes in observing conditions from night to night, slight shifts at the percent level in the zero-level occur between observation epochs. The observations were taken during a period of mediocre seeing, ranging from 1.5" to 2". Simultaneous I-Band (~630-1010 nm) observations obtained with the IAC80 telescope(de Ugarte Postigo and Castro-Tirado, priv. comm.), were used to calibrate count-rate to magnitudes.

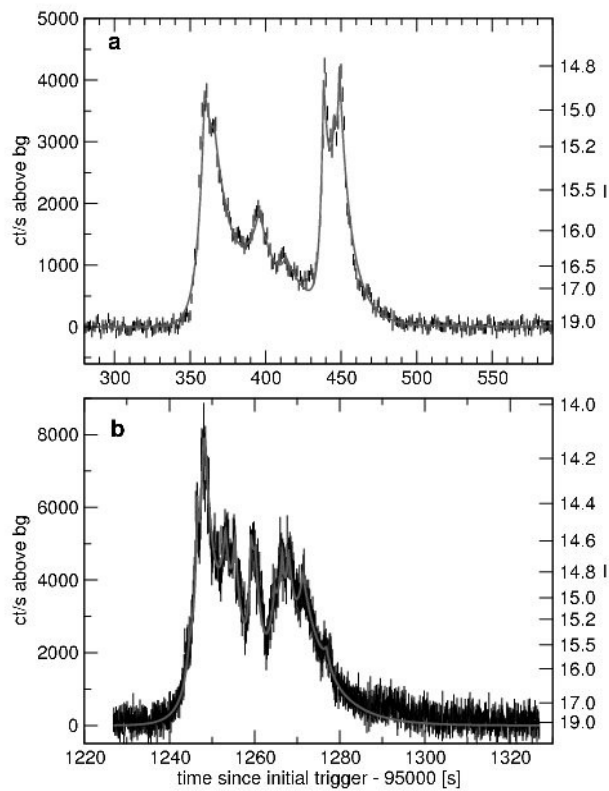

Figure 2: Light-curves of the two most prominent flares of the epoch 2007-06-11 shown in Fig.1. Panel a is binned to 1 s, panel b to 0.1 s resolution, to achieve reasonable signal-to-noise ratio while resolving the fastest variabilities detected in the data. The error bars are 1σ statistical errors. Extremely fast variability is clearly visible and well resolved. The solid line denotes a simple model, consisting of a superposition of several FRED shaped sub-flares. The FRED shape is modelled by an exponential rise with a short time-constant, followed by an exponential decay with a longer time-constant. The rise-time constants of the FREDs fitted to the light-curve in panel a range from 2.5 to 10 s, the decay time constants range from 3 s to 26 s. For panel b, the rise-time constants fitted to the light-curve range from 0.4 to 3 s, and the decay time-scales range from 0.3 to 7 s.

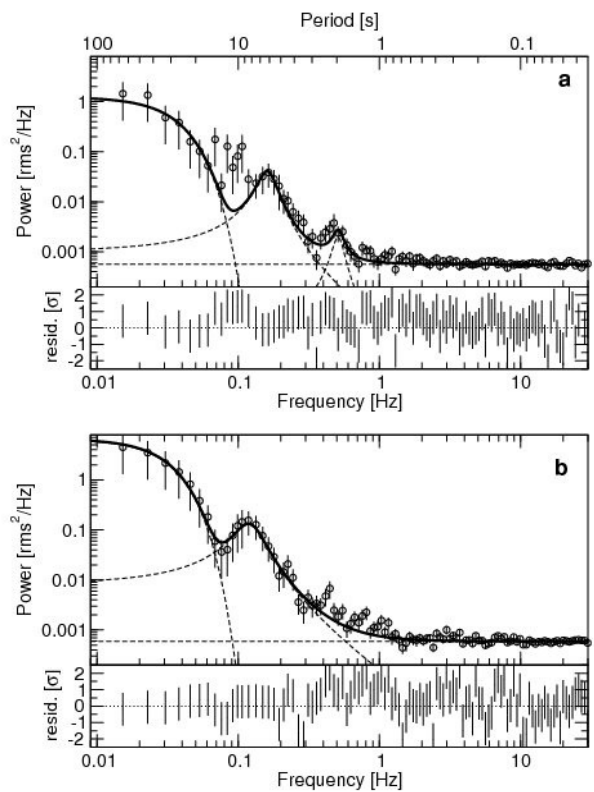

Figure 3: Power spectral density of the two prominent flares shown in panel a and b of Fig. 2. The power spectra are normalised such that their integral gives the squared rms fractional variability (therefore the power spectrum is in units of (rms)$^2$/Hz). The error bars are 1$\sigma$ statistical errors. The solid lines denote a best-fit model consisting of a Gaussian two Lorentzian and a constant component in panel a, and a Gaussian, one Lorentzian and a constant component in panel b. The dashed lines denote the individual components. For panel a, the reduced $\chi^2$ (sum of squared residuals / degrees of freedom) is 0.81, somewhat lower than desirable. The lower frequency Lorentzian in panel a has a central frequency of 0.16±0.007 Hz and a fwhm (full-width at half maximum) of 0.027±0.007 Hz, the higher frequency Lorentzian has a central frequency of 0.51±0.2 Hz and a fwhm of 0.05±0.02Hz. For panel b, the reduced $\chi^2$ is 0.97. The central frequency of the Lorentzian is 0.12±0.007 Hz, the fwhm is 0.028±0.007 Hz. Note that the lower frequency Lorentzian in panel a and b have central frequencies that overlap within the respective fwhms. Both power spectra stay flat above a frequency of a few Hertz up to 1 kHz (not plotted here), confirming the limit on the fastest variability-time-scales from the multi-component fits.